\documentclass[aps,onecolumn,superscriptaddress,floatfix]{revtex4}

\usepackage{bm}
\usepackage[colorlinks=true, citecolor=blue, urlcolor=blue]{hyperref}
\usepackage{epsf,graphics,graphicx}
\usepackage{amsmath}
\usepackage{latexsym}
\usepackage{amssymb,amsmath,graphicx}
\usepackage{amsfonts}
\usepackage[caption=false]{subfig}
\usepackage{epsfig,latexsym,graphicx}
\usepackage{amssymb}
\usepackage{graphicx}
\newcommand{\beq}{\begin{equation}}
\newcommand{\eeq}{\end{equation}}

\usepackage{epsfig}

\begin{document}
\title{Strata-based Quantification of Distributional Uncertainty in Socio-Economic Indicators: A Comparative Study of Indian States}
%	Comparison of Indian States over Three  Decades or,\\

	%or, \\
	%Entropic approach towards distributional uncertainty in socio-economic indicators: a comparative study of Indian States }
%\title{Understanding Distributional Uncertainty of Population, Literacy and Unemployment Rates within Indian States over Three Decades}
%\title{Distributional uncertainty in Socio-Economic Indicators: An  Entropic approach}

\author{Abhik Ghosh}
	\affiliation{Interdisciplinary Statistical Research Unit, Indian Statistical Institute, Kolkata 700108}
 \email{abhianik@gmail.com}
    \author{Olivia Mallick}
		\author{Souvik Chattopadhay}
	\author{Banasri Basu}
	\affiliation{Physics and Applied Mathematics Unit, Indian Statistical Institute, Kolkata 700108, India}
	\email{banasri@isical.ac.in}
\date{\today}

\begin{abstract}
This paper reports a comprehensive study of distributional uncertainty in  a few  socio-economic indicators 
across the  various states of India over  the years 2001-2011. 
We show that the DGB distribution, a typical rank order distribution, provide excellent fits to 
the  district-wise empirical  data for the population size, literacy rate (LR) and work participation rate (WPR) 
within every states in India, through its two distributional parameters. 
Moreover, taking resort to  the entropy formulation of the DGB distribution, a proposed uncertainty percentage (UP)
unveils the dynamics  of the uncertainty of LR and WPR  in all states of India. 
We have also commented on the changes in the estimated parameters and the UP values from the years 2001 to 2011. 
Additionally, a gender based analysis of the distribution of these important socio-economic variables 
within different states of  India has also been discussed. 
Interestingly, it has been observed that, although the distributions of the numbers of literate and working people 
has a direct (linear) correspondence with that of the population size, 
the literacy and work-participation rates are distributed independently of the population distributions.
\end{abstract}
\maketitle


\section{Introduction}

Socio-economic development of our world is majorly driven by people and their governance. 
The habitable area of the world is divided into distinct units, namely countries or dependent territories, 
governed  by independent administrative bodies of different nature. 
These countries are further subdivided (first tier division) into smaller units, sometimes called states or provinces, 
for the purpose of internal governance, management and better policy making.  
For the  administrative purpose and the ease of working in a local manner, 
these states or provinces are further subdivided  (second tier division) into more smaller strata called counties or districts.   
Although further subdivisions are seen in some countries, most large countries have major administrative divisions up to the second tier. 
For example, USA is divided into states (first-tier division) which, in turn, are divided into counties (second-tier division). 
China is divided into provinces and direct-controlled municipalities (first level), which are split into prefectures (second level), 
and prefectures are divided into counties (third level). 
India is divided into states (first tier) and  subsequently in districts (second-tier).  
Although there is wide heterogeneity among the administrative units (states or districts) in a country,
these  local administrative units can be seen as socially constructed strata, 
which serve as spatial scenarios for social and economic processes \cite{lopez2008}. 
Such administrative divisions are often regions of a country that are granted a certain degree of autonomy 
and manage themselves through local governments. 
% whereas some others have smaller subdivisions according to their requirements; the structure and size of administrative units is a trademark  of a country’s internal organization.  
In a growing number of developing countries, there is an attempt to proliferate sub-national administrative divisions for decentralization 
of  local governments \cite{grossman2014} such that the reorganization  may yield economic benefit for the cities \cite{ma2005}. 
Even in the absence of large administrative and political reforms, administrative units are constantly being created, 
destroyed, merged or split, demonstrating the fact that the sub-divisional strata within the countries undergo evolution.

The population distribution that %which
 does not distribute randomly \cite{eeckhout2004,kgbb,ag2019}  over different regions of a country 
is one of the key factors of the high degree of diversity and complexity \cite{christenson1980} 
of the internal regional  administration of various countries and territories.  
In this context, the study and understanding of the geographical distribution of the population within
a given country or region becomes relevant, as it is a necessary step for the development of theories 
that could accurately describe the evolution of human agglomerations \cite{batty2008,battym2013}. 
However, almost all studies regarding population distribution focus on city populations 
(see, for instance, \cite{kgbb,ag2019,soo2005,makse2008,holmes2010,jia2010,rybski2011,douady2011,masucci2013}), 
whereas the literature regarding the population distribution in administrative divisions  is scarce \cite{mansilla2009, betten2013}.

Moreover, there are studies predicting strong correlation of the population of any city with many of the characterizing features of the inhabitants of the corresponding city: %features  which show that a city's population strongly correlates with many of the features of the city’s inhabitants: 
mean income, number of registered patents per capita, criminality rates, land value and rent prices \cite{christenson1980,batty2008,ag2019}. 
In this respect, it is important to study the correlations of various socio-economic variables with the population distribution 
of the smaller strata of a country. In particular, to frame better policies for the whole country as well as for these subdivisions, 
it is necessary to understand, in detail, the distribution and associated uncertainty or inequalities of 
different socio-economic indicators within and between these strata along with their relations with the populations. 
These would then help us to classify different strata (subdivisions) within a country into similar groups in terms of the 
policy requirements for their developments where the groups are formed based on their underlying difference 
in terms of the socio-economic indicators. Is there any particular stratum within the country that requires special attention
than the rest or all the strata are equal in terms of their socio-economic characteristics?
In an attempt to answer such questions, in this paper, 
we will propose and discuss an innovative way of quantifying the distributional uncertainty of the socio-economic factors
within different strata (first tier subdivisions) of a country based on their rank-size pattern across the second-tier subdivisions. 

To study the uncertainty characteristics of a given  socio-economic factor within a strata (first-tier subdivision) of a country, 
we first fit an appropriate distribution to the rank-size values of this factor across all second-tier subdivisions within the strata.
The well known Zipf's law, alternatively known as power law or Pareto distribution \cite{zipf}%(**ref**), 
%a well known distribution, 
has been successfully used to model such rank-size data 
on socio-economic factors in its upper range of values,  but observed to fail in the lower end of the rank-size distribution \cite{lan2000, newman2005}. %(***\ref**). 
Very recently,  a universal framework through rank-order distribution  has been developed to 
successfully model a wide-range of rank-size data from various areas in arts and science \cite{Martinez-Mekler09, ag2019, Ausloos/Cerqueti:2016, Alvarez-Martinez/etc:2011, Alvarez-Martinez/etc:2014, Oscar2017, Alvarez-Martinez/etc:2018}  %(**ref**),
  where a two-parameter  discrete generalized beta (DGB) distribution is used instead of the one-parameter Pareto distribution. 
The DGB distribution is also shown to provide excellent fit to the rank-size data on various socio-economic parameters 
within a country and between different countries across the world \cite{ag2019}. 
Remarkably, this DGB distribution can indeed be obtained as an appropriate maximum-entropy distribution \cite{agmaxent}
and hence its entropy can be used to quantify the  maximum uncertainty present in the empirical data distribution. 
We will follow this idea to develop an uncertainty measure based on the entropy value of the fitted DGB distribution
to the empirical data on the given socio-economic factor across all second-tier subdivisions within the strata.

Our proposed uncertainty measure, which we refer to as the \textit{Uncertainty Percentage (UP)}, 
will indicate the distribution of the factor under study  
across different parts of the given strata; it will hence take the maximum value of 100\% 
if the factor value (socio-economic condition) is the same across all parts of the strata, i.e., in all second-tier subdivision within it,
indicating the equally distributed scenario (least inequality). On the other hand, our measure UP will take the minimum value of 0\%
if the inequality within the strata is extreme in that only one second-tier subdivision has a non-zero value of the factor under study
(socio-economic condition is as desired in only one subdivision)! Any value of UP in between these two extrema
will indicate how the targeted factor is distributed within different parts (subdivisions) of the strata
with higher values indicating more uniformity (less inequality).
Once this uncertainty measure is computed for all the strata within a country, they can be classified accordingly 
and can be used for their further in-depth comparisons.
For example, the fitted DGB distribution of different socio-economic factor 
can be linked by established appropriate relationship between the corresponding parameter estimates which,
along with our UP measure, provide further insights to characterize and classify different strata within a country for better policy making. 

%Additionally, we will also discuss how a multivariate regression model can be used to link the fitted DGB distribution 
%of any socio-economic factor with the population distribution within the strata. 
%the resulting relationship and the associated regression coefficients can also be used,
%along with our UP measure, to characterize and classify different strata within a country for better policy making. 

Further, by using our proposed idea, we will analyze uncertainty among the Indian states (first-tier strata)
based on the census data for the years 2011 and 2001. 
In this paper, we focus on the basic three socio-economic entities, namely, population, literacy and working population.
Like Population, education is a key element in any society that removes inequality from society, impacts the growth of employment and
improves a country’s gross national product. %Literacy is the basic building block and a crucial factor in the development of education in society. 
On the other hand, the distributional inequality of unemployment rates (that is given by the size of the non-working population)  plays an important role for structuring the economic development. 
These three factors, together controls, to a large extent, the development of the human resources in any society. 
For a vast and diverse country like India, which has so many states with very different  cultures, 
it is very important to analyses the levels of literacy and employment across various states of India and union territories (UTs) and investigate their underlying distributional uncertainty or inequality. 
We would like to apply our proposed idea \cite{ag2019, agmaxent} of fitting  the DGB distribution to the empirical data on the said three factors for all the districts of each Indian states. In a subsequent analysis we measure the  inequality across the second-tier subdivisions (districts) within a state via our proposed UP measure. 
The correlation of the distribution of literacy and employment rate within each state with its population distribution 
will be investigated by linking the parameters of the corresponding fitted DGB distributions.
Additionally, to strengthen our proposition we have also performed a gender based analysis for all these socio-economic indicators. 
Finally, our temporal analysis provide an overall picture  about the uncertainty or uniformity of the Indian states 
in terms of population, literacy and employment rates, their interrelations and changes from the year 2001 to 2011.

\section{Methodology}

\subsection{Modelling Empirical Data by DGB distribution}
\label{SEC:ROD}

%Let us first summarize the mathematics of the rank order distribution namely the discrete generalized beta distribution(DGBD). Among the many theories to explain city sizes, the growth dynamics and the distribution of the population across urban agglomeration the Pareto distribution have been most popular. Pareto distribution describes the negative relationship between the logarithm population size and logarithm of city rank. The mathematical form of this distribution  is
%%\begin{equation}
%%f_P(r) = A\cdot\frac{1}{r^\nu}, ~~~~r=1, \ldots, N,
%%\label{EQ:Pareto}
%%\end{equation} 
%where $r_i$ is the rank of the $i$-th item (or district) having size $n_i$ and $(r-1 ,\ldots,r_N)$ is a permutation of $(1  \ldots,N)$. $\nu>0$ is the model parameter, $A$ is the normalizing constant.As already noted in our earlier paper {REF:} this Pareto law provides good fit to the data only at low rank (large sizes) because of the fact the empirical data distribution often have an inflection point which is not captured by the single power law.

Let us consider a strata (e.g., state of a country) having $N$ second-tier subdivisions (e.g., districts).
Suppose that a socio-economic factor under study  (e.g., population, literacy, work participation rate, etc.) takes the value $x_i$  
at the $i$-th subdivision for $i=1, \ldots, N$.
We arrange these data in decreasing order of ``importance" (size) and denote the rank of the $i$-th subdivision as $r_i$ for $i=1, \ldots, N$.
We model these rank-size data $\{(r_i, x_i): i=1, \ldots, N \}$ by the discrete generalized beta (DGB) distribution 
having probability mass function
\begin{equation}
f_{(a,b)}(r) = A\frac{(N+1-r)^b}{r^a}, ~~~~r=1, \ldots, N.
\label{EQ:RO}
\end{equation}  
Here, in (\ref{EQ:RO}), $a, b$ are two real valued model parameters characterizing the underlying distributional structure 
and  $A$ is the normalizing constant depending on $(a, b)$ ensuring that $\sum_{r=1}^N f_{(a,b)}(r) =1$.
Note that, for the choice $b=0$, the probability mass function in (\ref{EQ:RO}) simplifies to that of the Pareto distribution.
With different values of the two parameters $(a, b)$, the class of DGB distribution in (\ref{EQ:RO}) allows us to model 
a wide range of rank-size distributions having an inflection point \cite{ag2019}   and
can successfully model different types of socio-economic factors \cite{Martinez-Mekler09, ag2019, Ausloos/Cerqueti:2016, Alvarez-Martinez/etc:2011, Alvarez-Martinez/etc:2014, Oscar2017, Alvarez-Martinez/etc:2018}%(**ref**).

Given empirical rank-size data $\{(r_i, x_i): i=1, \ldots, N \}$ of any socio-economic factor, 
we can fit the DGB distribution by estimating the model parameters $(a, b)$ by maximizing the likelihood of the observed data,
given by
\begin{equation}
L(a, b) = \prod_{i=1}^N f_{(a,b)}(r_i)^{{x}_i} = \prod_{i=1}^N \frac{(N+1-r_i)^{b{x}_i}}{r_i^{a{x}_i}}A^{{x}_i}.
\label{EQ:likelihood}
\end{equation}
This maximum likelihood estimator is known to be asymptotically most efficient and also enjoys several other optimality properties.
However, we need to use numerical optimization method to compute these estimates since the likelihood function in (\ref{EQ:likelihood})
does not posses an explicit form for its maximizer.
Once we get the maximum likelihood estimator (MLE), $(\widehat{a}, \widehat{b})$, of the model parameters $(a, b)$
we can then verify if the fitted DGB model is a good fit to the empirical data or not. 
For this purpose, we  use the overall error in model fit, computed in terms of the Kolmogorov-Smirnov (KS) measure 
between the observed and the predicted cumulative rank-sizes. 
Denoting the predicted size of rank $r_i$ as $p_i = \left(\sum_{i=1}^N x_i\right)f_{(\widehat{a}, \widehat{b})}(r_i)$ 
for all $i=1, \ldots, N$, we define the KS goodness-of-fit measure as 
\begin{eqnarray}
KS = \max_{1\leq i \leq N} \left|\left(\sum_{j: r_j\leq r_i} p_j\right) - \left(\sum_{j: r_j\leq r_i} x_j\right)\right|. 
\label{EQ:KS}
\end{eqnarray}
%As the fitted DGB distribution gives predictions close to the empirically observed rank-sizes,
%the KS values will be lower indicating better model fit.
The lower KS values indicate a better fitment of the model, i.e. the theoretically estimated fitted DGB distribution is very close to the empirically observed rank-sizes when the KS values are nearly vanishing. Zero KS value indicates perfect fit with no error for the observed data.

\subsection{ Quantification of Distributional Uncertainty}

Once a DGB distribution is found to be well fitted for a socio-economic indicator within a strata, 
we propose to quantify the underlying distributional uncertainty by its \textit{entropy}. 
Entropy is a widely used measure of disorder within any physical system.
Although the idea of entropy was originally used in Thermodynamics long back, 
its major use in information science and allied disciplines were started after Shannon's groundbreaking  work \cite{shanon1, shanon2} %(**ref**)
on mathematical formulation of entropy in an information channel and Jayne's Maximum entropy principle \cite{jaynes}.  
Subsequently, the idea of entropy and maximum entropy distribution has been widely used to analyze 
and assess the uncertainty in several geographical or socioeconomic variables \cite{kapurbook}  %(***ref***).

Here we also consider the DGB distribution, which is shown to be a maximum entropy rank-order distribution
under appropriate utility constraints \cite{agmaxent} %(**ref**).
 For this DGB distribution having probability as given in (\ref{EQ:RO}),
its Shannon entropy is given by 
\begin{eqnarray}
\label{entropy1}
S_N(a,b) &=& -\sum_{r=1}^N f_{(a,b)}(r) \log f_{(a,b)}(r)
%\\&&
= -\log A 
%\nonumber\\&& ~~
-A\sum_{r=1}^N \frac{(N+1-R)^b}{r^a} [b \log (N+1-r)-a \log r].
%\nonumber
\end{eqnarray}
This entropy value $S(a, b)$ then gives the maximum amount of uncertainty (entropy) 
lying within the underlying rank-order distribution for the given utility constraints
which are characterized  by the model parameters $(a, b)$. 
Therefore, given the empirical rank-size data, once we obtain the estimated parameter value  $(\widehat{a}, \widehat{b})$, 
we can estimate the (maximum) amount of uncertainty in these data by the entropy of the fitted DGB distribution,
i.e., by  $\widehat{S}_N = S_N(\widehat{a}, \widehat{b})$.

However, the estimated entropy value $\widehat{S}_N$ can vary from 0 to $\log(N)$
so it cannot be used to compare the uncertainty present in two distributions having different values of $N$.
Since the strata may often have different number ($N$) of subdivisions, we cannot compare them by just using $\widehat{S}_N$ 
as a measure of uncertainty. Standardizing by the range, we then define our proposed \textit{uncertainty percentage (UP)} measure as
the proportion of entropy estimated from the fitted DGB distribution with respect to its maximum possible value, 
i.e., 
\begin{equation}
UP = \frac{ S_N(\widehat{a}, \widehat{b})}{\log N}\times 100 \%.
\end{equation}
Note that, clearly the UP measure takes the value 100\% if $S_N(\widehat{a}, \widehat{b})=\log(N)$
which holds if and only if $\widehat{a}=0= \widehat{b}$ indicating that 
the best-fitted rank-sizes are uniformly distributed over the subdivisions within the given strata;
in other words, there is no inequality of the distribution of the targeted socio-economic indicator within different parts of the strata. 
On the other hand, the UP measure will be zero if $S_N(\widehat{a}, \widehat{b})=0$
indicating the extremely dispersed distribution of the rank-sizes within the strata 
with only one subdivision having non-zero value of the targeted indicator. 
Therefore, the UP can be used to quantify the distributional uncertainty of the spread of the distribution 
of any socio-economic indicator within a strata measuring its $closeness$ to the optimum case of uniformity. 
Since the UP measure takes the value from 0 to 100\%, irrespective of the value of $N$, 
it can then also be used to compare the uncertainty present within different strata as well as for different indicators.

\subsection{Inter-relations between the distributions of two socio-economic variables }

%*** TO remove or to keep? ***

In any human agglomeration unit, most of the socio-economic indicators are highly correlated with each other. In particular, we have noted that the population distribution 
has a huge impact  on the distribution of various socio-economic factors that had been explored in the literature
via regression within the power law structure. %(**ref**). 
However, in this paper, we have used  the alternative DGB rank-order distribution that is noted %seen 
to yield better fitment of the model to the empirical data. %better model fit to the empirical data.
Being consistent with the fitted DGB distributions, we here propose to investigate the relationship between the distributions of any socio-economic indicator with the population distribution  within different strata. 
To be more specific, our approach focus more on the inter-relation between the distributional structure and uncertainty of the socio-economic variables rather than the corresponding values of the indicators.

To investigate the relationship between the DGB distributions fitted to two variables within a strata, 
let us recall that the DGB distributions are characterized by its two model parameters $(a, b)$.
Suppose that there are $T$ strata and, for the $j$-th stratum, 
the MLE of the parameters $(a,b)$ corresponding to the fitted population distribution is $(\widehat{a}_{Pj}, \widehat{b}_{Pj}) $ 
and the same for any targeted socio-economic indicator $Y$ (e.g., literacy or unemployment rate) is  $(\widehat{a}_{Yj}, \widehat{b}_{Yj})$,
$j=1, \ldots, T$. Then to study the relationship between the best-fitted DGB distributions to the targeted indicator $Y$ and the population
across the  $T$ strata, it is enough to investigate the relationship between the corresponding parameters 
$(\widehat{a}_{Pj}, \widehat{b}_{Pj}) $ and   $(\widehat{a}_{Yj}, \widehat{b}_{Yj})$ for $j=1, \ldots, T$.
The first attempt in this regard should be the examination of the (Pearson) bivariate correlations among these estimated parameters across strata.
If this correlation is significantly higher (in absolute value), it indicates the  linearity of their relationship 
which we may then further explore via appropriate multivariate regression models. 
%as given by 
%\begin{eqnarray}
%(\widehat{a}_{Yj}, \widehat{b}_{Yj})^T  = B(\widehat{a}_{Pj}, \widehat{b}_{Pj})^T + (\epsilon_{1j}, \epsilon_{2j})^T, 
%~~~~j=1, \ldots, T.
%\label{EQ:reg}
%\end{eqnarray}
%Here, $\epsilon_{1j}$ and $\epsilon_{2j}$ are independent and identically distributed random error components having mean zero and constant variances,
%and $B$ is a $3\times 2$ matrix of regression coefficient as
%$
%B= \begin{bmatrix}
%\begin{array}{ccc}
%\beta_{a,0} & \beta_{a,a} & \beta_{a,b}\\
%\beta_{b,0} & \beta_{b,a} & \beta_{b,b}
%\end{array}
%\end{bmatrix}.
%$
%It is to be noted that these regression coefficients have  the interpretation similar to that of a simple linear regression model
%indicating the change in the distributional parameters of the targeted indicators caused by change in one unit of %change in 
%the DGB parameters of the population distribution. 
%These can be estimated by standard least square method or by the maximum likelihood estimator assuming normality of the error components;
%both the approaches  lead to the same and most efficient estimator of the coefficient matrix $B$
%which we can use to infer about the relationship between the two fitted DGB distributions across strata. 
%However, we also need to check the accuracy of the linear relationship (\ref{EQ:reg}) using the resulting coefficient of determination ($R^2$)
%and use this approach only if the value of $R^2$ is close enough to 1, e.g., if $R^2 \geq 0.8$ or so depending on the context.
The same can be studied using the UP values and also for any two socio-economic indicators (other than the population).

However, if there is no linear relationship between the fitted parameters or the UP values corresponding to two indicators,
one should investigate their relation via appropriate non-parametric correlation coefficients (e.g., rank correlation).
If this non-parametric correlation is found to be significant, then one may proceed for further analyses
of their non-linear relationship using advanced statistical tools, which  will not be considered in the present paper.

\section{Results: Understanding Uncertainty among Indian States } %or, Understanding Uniformity/ Non-uniformity among Indian States}
\label{SEC:RES}

\subsection{Data Description and Overall Statistics}
\label{SEC:data}

We consider the data from Indian census which is conducted in every 10 years to capture 
a detailed picture of demographic, economic and social conditions of all persons in the country pertaining to that specific time.
The raw census data for the years 2011 and 2001 are obtained from the \textit{Primary Census data and Digital library} (www.censusindia.gov.in). 
Following the stratified administrative structure of India, 
we use the districts as smaller second-tier units within each state (first-tier strata) of India. 
%India is a federal union, at present comprising of 28 states and 8 union territories (UTs), for a total of 36 entities. 
According to the final census in 2011, there were  29 states and 7 UTs in India, which consist of the pool of strata in our analysis;
for simplicity we will also refer the UTs as the \textit{State}.
It is important to note that some states, created at a later time, were not present in the earlier round of census (at the year 2001).
The total number of districts within all the states of India was 640 and 593 in the years 2011 and 2001, respectively. 
Additionally, we will also use these census data separately for male and female populations 
for further gender-based uncertainty analysis in Section \ref{SEC:gender_results}.

\begin{figure}[!b]
	\centering
	%\subfloat[2011]{
	\includegraphics[width=0.7\textwidth]{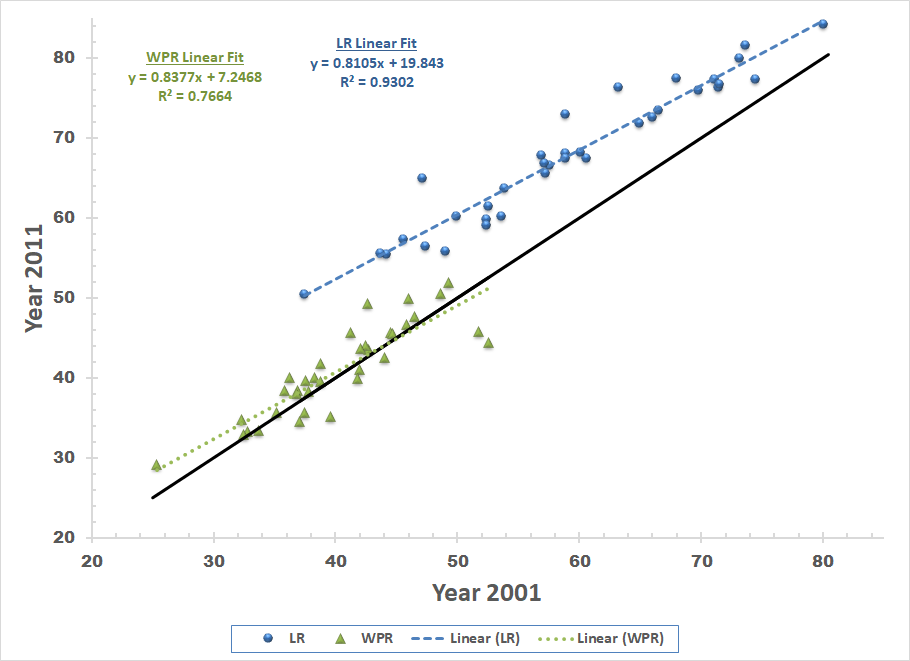}
	%\label{FIG:loglogCumHaz_Veteran}}
	%~ %--------------------------------------------------------------------
	%\subfloat[2001]{
	%\includegraphics[width=0.3\textwidth]{boxplot01.png}
	%	\label{FIG:loglogCumHaz_Veteran}}
	%~	%---------------------------------------------------------------------------
	\caption{Plot of LR and WPR for different states in the years 2001 and 2011, along with the corresponding linear fits. 
	The black solid line represent the $y=x$ line (linear growth)}
	\label{FIG:LR_WPR}
\end{figure}

As mentioned earlier, in this paper, we focus on two important socio-economic factors, 
namely literacy and work participation rates, along with the population of any unit under consideration. 
For literacy computation, a person is defined to be literate  if (s)he is aged more than 7 years 
and can both read and write with understanding in any language.
On the other hand, a person who has worked in any economically productive activity for more than 6 months 
during the last one year preceding the data collection date is considered to be a worker (main or marginal).  
We analyze the data on the corresponding variables (in rate) defined as follows:  
\begin{eqnarray}
\mbox{Literacy rate (LR)} &=& \frac{\mbox{number of literate people in a district}}{\mbox{total population in that district}}\times 100,
\nonumber\\
\mbox{Work Participation Rate (WPR)} &=& \frac{\mbox{number of Workers in a district}}{\mbox{total Population in that district}}\times100.
\nonumber
\end{eqnarray}
%the Literacy rate (LR)=(number of literate people in a district)/(total population in that district)*100
%and the Work Participation Rate (WPR)=(number of Workers in a district)/(total Population in that district)*100;
A comprehensive summary of state-wise population (P), LR and WPR statistics (in \%) are reported in Supplementary Table \ref{TAB:Primary_data}
(Appendix \ref{SEC:App}) for the years 2011 and 2001. It can easily be seen that the population of each state has increased over the years (linearly).
%but the changes in WPR is rather variables. 
To understand the changes in LR and WPR more clearly,  we have plotted their values in the years 2001 versus in the year 2011 
in Figure \ref{FIG:LR_WPR}, along with the corresponding linear fit.  
According to 2011 census India has literacy rate (LR) around 84\% to 50\%, among which Kerala is highest and Bihar is lowest;
but, the WPR for different states of India varies in between 35\% to 50\% only. 
Further, the LR has significantly increased in all the states whereas the changes in WPR is rather variables. 	
Except for Mizoram and Kerala, all states has an increase of more than 5\% in LR, with as many as 9 states having an increase of more than 10\%.
On the other hand, 10 states have no significant changes (only less than 1\%) in the WPR.
The UT of  Dadra and Nagar Haveli has the highest increase in LR but maximum decrease in WPR.
Nagaland is seen to have the highest increase of 6.64\% in the WPR. 
Among larger states, Kerala, Himachal Pradesh, Assam, Jharkhand and Odisha has shown some amount of increase ($>2\%$) in WPR.
Overall Tripura seems to have performed best considering both LR and WPR.
On an average, there have been sublinear growths in both LR and WPR among the Indian states and UTs.

%**********************************\\
%**** We need to add here the state-wise overall literacy \% and unemployment rate (in \%) and 
%their changes over the three decades for some basic overall comparisons *** then we go on to the deeper analysis of their distributions **
%*** we can either add some table or graph as you think appropriate ****

\subsection{Modelling Population, Literacy and Work participation rates within each States}
\label{SEC:RO-results}

\noindent
\textbf{\small Fitting the DGB rank-order Distributions within each States}: 

We now discuss the results obtained by the application of the DGB distribution for different socioeconomic variables (Population, LR and WPR)
for various states of India using the district-wise data. 
%In our previous paper we have shown the superiority of the DGBD over the Pareto model for different kind of city size distribution for different economics and geographies across the world. 
%As already mentioned, here we consider districts as the smallest administrative unit for a state and we have fitted the DGB, a typical rank-order (RO) distribution for all the states for the variables population (P), number of literate persons (L) and number of employed persons or, workers (W). 
The states comprising of districts less than 5 are excluded from the analysis;
this excludes 6 UTs (all except Delhi) and three states (Goa, Sikkim and Tripura).  
The values of the estimated parameters ($a, b $) and the goodness-of-fit measures (KS) for all three socioeconomic variables for different states, 
are reported in Table \ref{TAB:DGB_results} for both the years 2011 and 2001.
As expected (see \cite{ag2019,agmaxent}), the DGB distribution, with the two distributional parameters $\widehat{a}$ and $\widehat{b}$, 
provide excellent fits  to the empirical data not only for Population size but also for LR and WPR with extremely small KS error.

Considering the distributional structures, for most states, both the parameters $\widehat{a}$ and $\widehat{b}$ 
tend to have positive values and are also nature-wise similar.
However, there has been significant variations among the estimated parameter values across different states for all the variables,
indicating differences in their distributions with the states (across districts).  
Particularly the UT of Delhi, the capital of India and one of the crowded states, has a negative $\widehat{a}$ value  $-0.157$ 
and a positive $\widehat{b}$ value 1.387 for the distribution of populations. 
On the other hand, for modelling population, the not so much populated state Mizoram  has a high positive value of
$\widehat{a}$ (0.999) than any other states but a negative  $\widehat{b}$ value. 
%Arunachal Pradesh has a very small $\widehat{a}$ value of 0.074.
While considering the distribution of LR, most of the states fits to the positive values for both $\widehat{a}$ and $\widehat{b}$,
except for Chandigarh, Delhi, Haryana, Mizoram which yield negative  $\widehat{a}$ value and positive $\widehat{b}$ value. 
In 2011, Kerala, the state with 84 percent literate people, corresponds to $\widehat{a}$ and $\widehat{b}$ values of 0.018 and 0.035, respectively. 
West Bengal having 67.4 percent literate people gives $\widehat{a}$ and $\widehat{b}$ values 0.043 and 0.127. 
%To an exception, for Delhi $\widehat{a}$ and $\widehat{b}$ values are given by -0.1739 and 1.4076., which inherently remain same as population size. 
For the distribution of WPR as well, some states have negative $\widehat{a}$ value indicating their difference from the majority of the states.
In general, disparity of LR among the states are more than that of WPR.
%******entropy values for LR and WPR are greater than entropy value for population for  a particular state.
%The detailed data is given in Table I and the results of our analysis are presented in Tables ********.

\begin{table}
	\caption{The parameter estimates and the KS measures for for different states along with total numbers (N) of districts }
	%	\resizebox{0.6\textwidth}{!}{ 
	\begin{tabular}{l|r|rrr|rrr|rrr}\hline
&\multicolumn{1}{|c|}{}  &\multicolumn{3}{|c|}{Population} &\multicolumn{3}{|c|}{LR} &\multicolumn{3}{|c}{WPR} 	\\
States   &	$N$	 &$\widehat{a}$ 	&$\widehat{b}$	&KS& 	$\widehat{a}$	&	$\widehat{b}$&	KS	& 	$\widehat{a}$	&	$\widehat{b}$&	KS\\
\hline\hline
\multicolumn{11}{c}{\underline{Year 2011}}\\		
		India	&	640	&	0.252	&	0.872	&	0.005	&	0.063	&	0.133	&	0.004	&	0.082	&	0.109	&	0.006	\\
		Andhra Pradesh	&	23	&	0.124	&	0.171	&	0.007	&	0.074	&	0.050	&	0.001	&	-0.009	&	0.109	&	0.001\\	
%		Andaman Nicobar	&	3 &-	&-		&-		&-		&-		&-		&-		&-		&-	\\
		Arunachal Pradesh	&	16	&	0.074	&	0.825	&	0.020	&	0.069	&	0.125	&	0.002	&	0.122	&	0.033	&	0.003\\	
		Assam	&	27	&	0.265	&	0.307	&	0.011	&	0.076	&	0.070	&	0.002	&	0.060	&	0.071	&	0.006\\	
		Bihar	&	38	&	0.143	&	0.534	&	0.003	&	0.058	&	0.086	&	0.006	&	0.064	&	0.056	&	0.003\\	
%		Chandigarh	&	1	&-	&-		&-		&-		&-		&-		&-		&-		&-	\\
		Chattishgarh	&	18	&	0.390	&	0.664	&	0.015	&	-0.022	&	0.285	&	0.005	&	0.044	&	0.059	&	0.002\\	
%		Dadra Nagar Haveli	&	1 &-	&-		&-		&-		&-		&-		&-		&-		&-	\\
%		Damman Diu	&	2	&-	&-		&-		&-		&-		&-		&-		&-		&-	\\
		Delhi	&	9	&	-0.157	&	1.387	&	0.026	&	0.015	&	0.033	&	0.001	&	0.101	&	0.022	&	0.006\\	
%		Goa	&	2&-	&-		&-		&-		&-		&-		&-		&-		&-	\\
		Gujarat	&	26	&	0.383	&	0.454	&	0.020	&	0.021	&	0.112	&	0.003	&	0.112	&	0.011	&	0.003\\	
		Haryana	&	21	&	0.140	&	0.199	&	0.006	&	-0.014	&	0.150	&	0.004	&	0.020	&	0.105	&	0.002\\	
		Himachal Pradesh	&	12	&	0.306	&	0.748	&	0.035	&	0.009	&	0.076	&	0.001	&	0.059	&	0.119	&	0.005\\	
		Jammu Kashmir	&	22	&	0.368	&	0.487	&	0.011	&	0.141	&	0.049	&	0.003	&	0.175	&	0.046	&	0.004\\	
		Jharkhand	&	24	&	0.271	&	0.394	&	0.013	&	0.060	&	0.107	&	0.003	&	0.052	&	0.111	&	0.004\\	
		Karnataka	&	30	&	0.689	&	0.033	&	0.031	&	0.047	&	0.122	&	0.002	&	0.044	&	0.032	&	0.002\\	
		Kerala	&	14	&	0.056	&	0.592	&	0.011	&	0.018	&	0.035	&	0.002	&	0.074	&	0.107	&	0.006\\	
%		Lakshadweep	&	1	&-	&-		&-		&-		&-		&-		&-		&-		&-	\\	
		Madyapradesh	&	50	&	0.197	&	0.293	&	0.005	&	0.016	&	0.159	&	0.005	&	0.048	&	0.098	&	0.001\\	
		Maharashtra	&	35	&	0.491	&	0.288	&	0.014	&	0.027	&	0.077	&	0.001	&	0.052	&	0.038	&	0.002\\	
		Manipur	&	9	&	0.126	&	0.623	&	0.030	&	0.038	&	0.103	&	0.003	&	0.054	&	0.059	&	0.001\\	
		Meghalayas	&	7	&	0.376	&	0.477	&	0.014	&	0.055	&	0.110	&	0.007	&	0.036	&	0.025	&	0.002\\	
		Mizoram	&	8	&	0.990	&	-0.119	&	0.020	&	-0.046	&	0.239	&	0.006	&	-0.008	&	0.184	&	0.003\\	
		Nagaland	&	11	&	0.241	&	0.563	&	0.012	&	0.050	&	0.171	&	0.004	&	0.107	&	0.109	&	0.005\\	
		Oddisha	&	30	&	0.239	&	0.488	&	0.007	&	0.027	&	0.230	&	0.005	&	0.035	&	0.167	&	0.008\\	
%		Pondicherry	&	4	&-	&-		&-		&-		&-		&-		&-		&-		&-	\\	
		Punjab	&	20	&	0.381	&	0.332	&	0.024	&	0.042	&	0.082	&	0.007	&	0.062	&	0.021	&	0.003\\	
		Rajasthan	&	33	&	0.367	&	0.211	&	0.012	&	0.062	&	0.066	&	0.003	&	0.063	&	0.049	&	0.001\\	
%		Sikkim	&	4	 &-	&-		&-		&-		&-		&-		&-		&-		&-	\\ 
		TamilNadu	&	32	&	0.175	&	0.477	&	0.009	&	0.042	&	0.050	&	0.001	&	0.035	&	0.079	&	0.001\\	
%		Tripura	&	4	&-	&-		&-		&-		&-		&-		&-		&-		&-	\\
		Uttarakhand	&	13	&	0.478	&	0.542	&	0.037	&	0.014	&	0.061	&	0.002	&	-0.001	&	0.191	&	0.006\\	
		Uttar Pradesh	&	71	&	0.147	&	0.394	&	0.018	&	0.030	&	0.116	&	0.002	&	0.079	&	0.026	&	0.003\\	
		West Bengal	&	19	&	0.255	&	0.364	&	0.006	&	0.043	&	0.127	&	0.002	&	0.051	&	0.024	&	0.001\\	
\hline\hline	
%\end{tabular}
%	\begin{tabular}{r|r|rrr|rrr|rrr}\hline
%		&\multicolumn{1}{|c|}{}  &\multicolumn{3}{|c|}{Population} &\multicolumn{3}{|c|}{LR} &\multicolumn{3}{|c|}{WPR} 	\\\									
%		STATES   &	$N$	 &$\widehat{a}$ 	&$\widehat{b}$	&KS& 	$\widehat{a}$	&	$\widehat{b}$&	KS	& 	$\widehat{a}$	&	$\widehat{b}$&	KS\\\hline
\multicolumn{11}{c}{\underline{Year 2001}}\\		
		India	&	593	&	0.236	&	0.808	&	0.007	&	0.077	&	0.192	&	0.004	&	0.074	&	0.123	&	0.005	\\	
		Andhra Pradesh	&	23	&	0.125	&	0.148	&	0.009	&	0.103	&	0.077	&	0.003	&	-0.016	&	0.150	&	0.004\\		
%		Andaman Nicobar	&	2	&-	&-		&-		&-		&-		&-		&-		&-		&-		\\		
		Arunachal Pradesh	&	13	&	0.130	&	0.490	&	0.011	&	0.103	&	0.131	&	0.005	&	0.082	&	0.080	&	0.002\\		
		Assam	&	23	&	0.252	&	0.357	&	0.012	&	0.077	&	0.099	&	0.004	&	0.146	&	0.041	&	0.005		\\
		Bihar	&	37	&	0.152	&	0.461	&	0.004	&	0.102	&	0.140	&	0.006	&	0.081	&	0.081	&	0.006		\\
%		Chandigarh	&	1	&-	&-		&-		&-		&-		&-		&-		&-		&-		\\	
		Chattishgarh	&	16	&	0.456	&	0.248	&	0.017	&	-0.038	&	0.322	&	0.009	&	0.037	&	0.081	&	0.004\\		
%		Dadra Nagar Haveli	&	1	&-	&-		&-		&-		&-		&-		&-		&-		&-		\\		
%		Damman Diu	&	2   &-  &-		&-		&-		&-		&-		&-		&-		&-		\\
		Delhi	&	9	&	-0.094	&	1.163	&	0.021	&	0.003	&	0.051	&	0.001	&	0.027	&	0.077	&	0.003		\\
%		Goa	&	2	&-	&-		&-		&-		&-		&-		&-		&-		&-		\\
		Gujarat	&	25	&	0.360	&	0.419	&	0.018	&	0.019	&	0.168	&	0.003	&	0.051	&	0.065	&	0.002	\\	
		Haryana	&	19	&	0.263	&	0.168	&	0.006	&	0.031	&	0.086	&	0.003	&	0.018	&	0.112	&	0.003	\\	
		Himachal Pradesh	&	12	&	0.326	&	0.714	&	0.035	&	-0.002	&	0.109	&	0.002	&	0.114	&	0.033	&	0.003\\		
		Jammu Kashmir	&	14	&	0.166	&	0.766	&	0.019	&	0.182	&	0.077	&	0.005	&	0.166	&	0.083	&	0.017\\		
		Jharkhand	&	18	&	0.083	&	0.661	&	0.014	&	0.134	&	0.173	&	0.006	&	0.060	&	0.130	&	0.003\\		
		Karnataka	&	27	&	0.464	&	0.120	&	0.020	&	0.083	&	0.125	&	0.002	&	0.030	&	0.065	&	0.001\\		
		Kerala	&	14	&	0.021	&	0.596	&	0.012	&	0.021	&	0.048	&	0.002	&	0.085	&	0.104	&	0.005\\		
%		Lakshadweep	&	1	&-	&-		&-		&-		&-		&-		&-		&-		&-		\\		
		Madyapradesh	&	45	&	0.104	&	0.357	&	0.004	&	0.030	&	0.163	&	0.004	&	0.061	&	0.090	&	0.002	\\	
		Maharashtra	&	35	&	0.449	&	0.298	&	0.013	&	0.040	&	0.098	&	0.001	&	0.030	&	0.061	&	0.001	\\	
		Manipur	&	9	&	0.350	&	0.409	&	0.029	&	0.098	&	0.077	&	0.003	&	0.024	&	0.075	&	0.002\\		
		Meghalaya	&	7	&	0.380	&	0.527	&	0.018	&	0.202	&	0.025	&	0.005	&	0.034	&	0.075	&	0.001\\		
		Mizoram	&	8	&	1.116	&	-0.305	&	0.009	&	-0.001	&	0.203	&	0.003	&	0.081	&	0.098	&	0.002	\\	
		Nagaland	&	8	&	0.284	&	0.261	&	0.015	&	-0.005	&	0.353	&	0.006	&	0.044	&	0.178	&	0.008	\\	
		Oddisha	&	30	&	0.246	&	0.481	&	0.008	&	0.024	&	0.334	&	0.005	&	0.057	&	0.179	&	0.010	\\	
%		Pondicherry	&	4	&-	&-		&-		&-		&-		&-		&-		&-		&-		\\		
		Punjab	&	17	&	0.310	&	0.470	&	0.022	&	0.046	&	0.127	&	0.007	&	0.060	&	0.046	&	0.001	\\	
		Rajasthan	&	32	&	0.331	&	0.225	&	0.012	&	0.080	&	0.086	&	0.003	&	0.066	&	0.051	&	0.003 \\		
%		Sikkim	&	4	&-	&-		&-		&-		&-		&-		&-		&-		&-		\\	
		TamilNadu	&	30	&	0.176	&	0.495	&	0.008	&	0.059	&	0.056	&	0.001	&	0.055	&	0.108	&	0.001	\\	
%		Tripura	&	4	&	-	&	-	&	-	&	-	&	-	&	-	&	-	&	-	&	-	\\	
		Uttarakhand	&	13	&	0.364	&	0.552	&	0.023	&	0.040	&	0.081	&	0.003	&	0.007	&	0.208	&	0.004	\\	
		Uttar Pradesh	&	70	&	0.142	&	0.383	&	0.016	&	0.054	&	0.161	&	0.002	&	0.095	&	0.051	&	0.004	\\	
		West Bengal	&	18	&	0.312	&	0.395	&	0.010	&	0.063	&	0.177	&	0.002	&	0.077	&	0.042	&	0.002	\\	
		\hline
	\end{tabular}
	%}
	\label{TAB:DGB_results}
\end{table}

Next, if we compare the fitted distribution (their estimated parameters) for the two years 2011 and 2001,
they changes more for the smaller states and the states having significantly change in district numbers.
Among the states having mostly the same numbers of district in both 2001 and 2011, 
significant change in the values of the estimated parameters are observed for Bihar and meghalay while considering the LR distribution
but the others remains mostly the same. However, the distribution of WPR has significant changes in more states 
including Assam, Delhi, Gujarat, Mizoram and Tamil Nadu. 
%We will further use these estimated parameters to classify the states accordingly to the distributional structures of
%these three important variables later in Section \ref{SEC:classification}.

%\begin{figure}[!h]
%	\centering
	%\subfloat[2011]{
		%\includegraphics[width=0.3\textwidth]{boxplot11.png}
		%\label{FIG:loglogCumHaz_Veteran}}
	~ %--------------------------------------------------------------------
	%\subfloat[2001]{
	%\includegraphics[width=0.3\textwidth]{boxplot01.png}
%	\label{FIG:loglogCumHaz_Veteran}}
~	%---------------------------------------------------------------------------
%	\subfloat[1991]{
	%\includegraphics[width=0.3\textwidth]{boxplot91.png}
%	\label{FIG:loglogCumHaz_Veteran}}
 %--------------------------------------------------------------------
	%\caption{Plot of KS RO and KS Pareto for three variables for 2011, 2001, 1991 }
	%\label{FIG:box-plot}
%\end{figure}

%\subsection{Distributional Uncertainty Analysis: How different are the States? }
\noindent
\textbf{\small Distributional Uncertainty Analysis: }
\label{SEC:UP_results} 
 
\begin{figure}[!h]
\centering
\subfloat[Year 2011]{
	\includegraphics[width=0.33\textwidth]{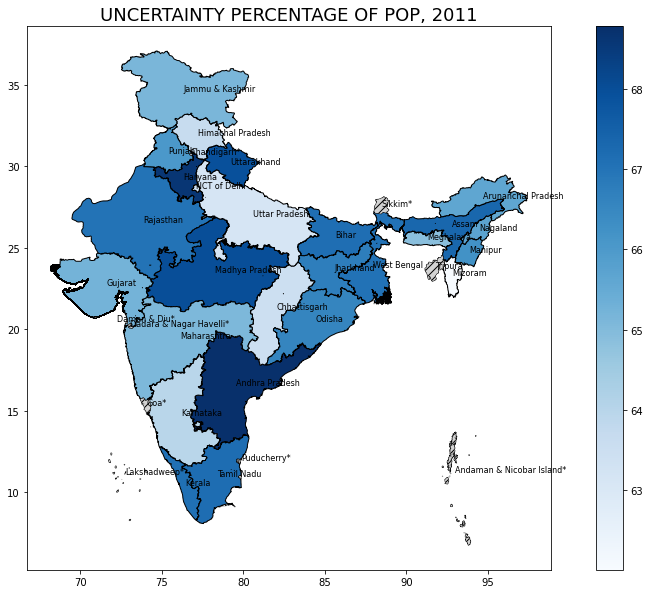}
	\includegraphics[width=0.33\textwidth]{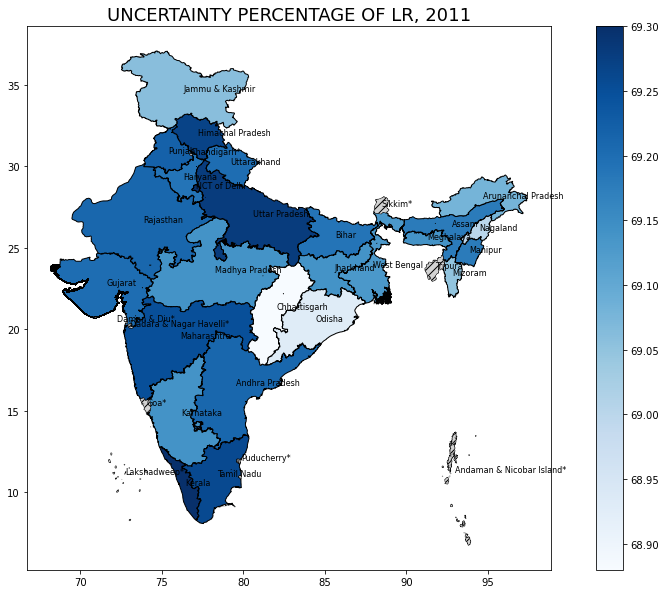}
	\includegraphics[width=0.33\textwidth]{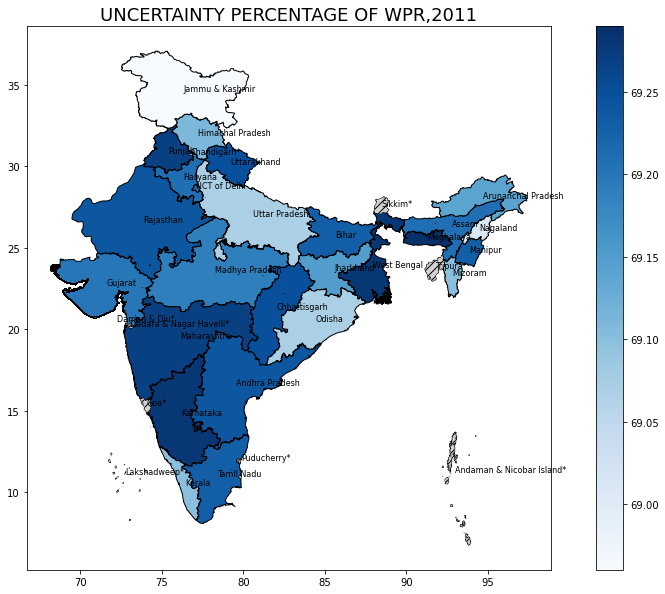}
	\label{FIG:loglogCumHaz_Veteran}}
	\\ %--------------------------------------------------------------------
\subfloat[Year 2001]{
	\includegraphics[width=0.33\textwidth]{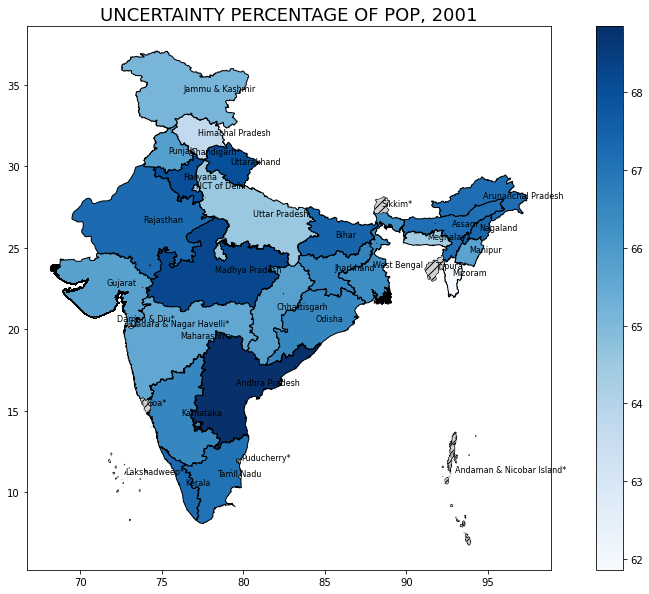}
	\includegraphics[width=0.33\textwidth]{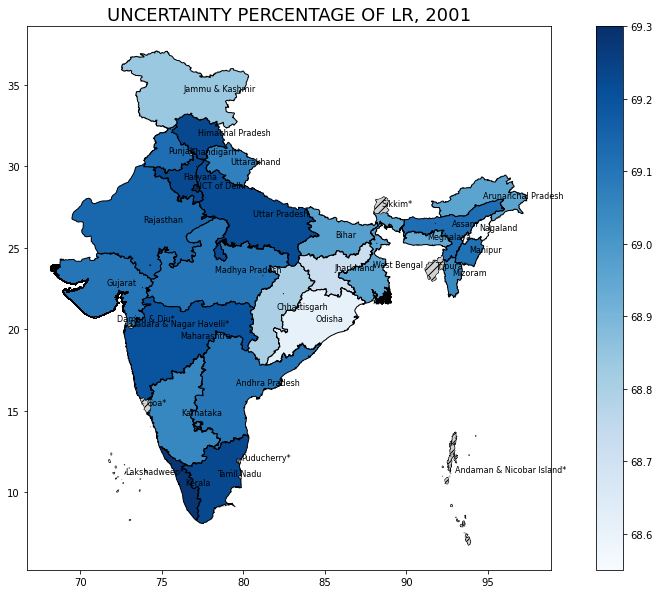}
	\includegraphics[width=0.33\textwidth]{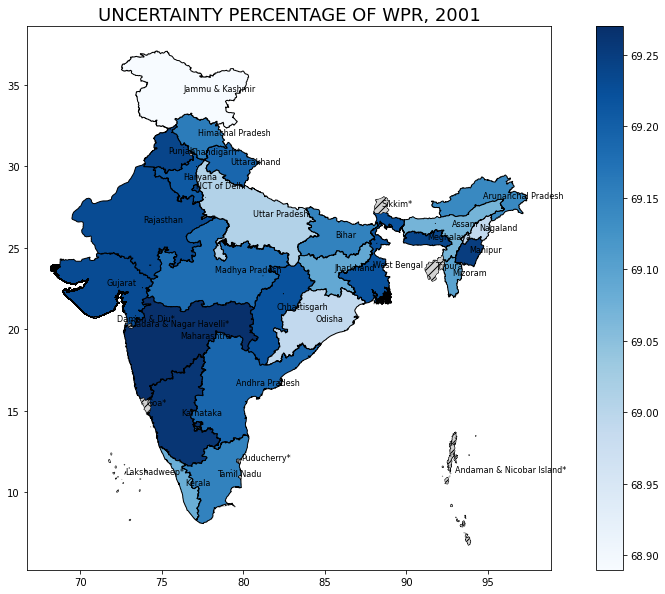}
	\label{FIG:loglogCumHaz_SmallCell}}
	%---------------------------------------------------------------------------
\caption{Plots of UP for Populations, LR and WPR for different states of India in 2011 and 2001}
\label{FIG:UP}
\end{figure}

We have revealed that the DGB distribution can completely characterize the population size, LR and WPR distributions 
through two $sensors$ or $drivers$ modelled by two parameters ($\widehat{a}$ , $\widehat{b}$). 
The predicted distribution fits very well with the observed rank-size data of all the districts of India 
as well as the data of the districts within the states. 
We now find the $entropy$ (S) values of all the considered distributions and 
%consequently find the relation of  $S$ with the distributional parameters $(a,b) $. 
% It is also noted that the estimated entropy value or our approach relates the estimated parameters to the entropy of the distribution namely DGBD. 
%Our present purpose is to demonstrate the applicability of Shannon entropy to social and economic life of a human settlement, by applying the DGBD distribution to different socio-economic variables. For this analysis we 
take resort to the proposed entity $Uncertainty$ $Percentage$ (UP) by standardizing the $entropy$ $S$ values of 
the three variables of different Indian province. 
The resulting values of UP are plotted in Figure \ref{FIG:UP} for both the years 2011 and 2001.
%Next we discuss some salient features of the results in the context of equitability index. ****** 

One can see from Figure \ref{FIG:UP} that the UP values changes mostly for the population and least for the WPR. 
So, there has been no significant changes in the distributional uncertainty of the work participation rates in all states on India.
A little significant increase of 0.14\% only is observed for the UP of WPR in Assam whereas it changes less that 0.1\% in all other states.
The changes in the UP values for LR is also less than 0.5\% indicating no significant changes in 
the distributional uniformity of literacy rates in all the states in India.   
For the population, however, maximum increase (of 0.72\%) in the UP values is observed in  West Bengal 
indicating greater uniformity of its distribution within the states. 
On the other hand, in Karnataka and Chattisgarh, the UP values of population distribution decreases more than 2\% 
indicating significant relocations of people within these states (mostly from rural areas to major cities).

In absolute term, Andhra Pradesh, Madhya Pradesh and Haryana  has maximum UP values and hence more uniform population distribution, 
whereas those in Mizoram and Uttarakhand are more sparse. In terms of literacy rate, however, Delhi, Kerala and Uttarakhand 
are most evenly distributed states having highest UP values and Chhattisgarh and  Odisha  have most uneven distribution of LR (lowest UP values).
Finally, Karnataka, West Bengal and Meghalaya have highest UP values for WPR indicating most equally distributed work-participation rates 
among their districts, whereas Jammu and Kashmir, Nagaland and Odisha have most uneven workforce within the states.
\\\\\\
 
%\subsection{Relationship of the Literacy and Work participation rate Distribution with Populations across the States}
%\label{SEC:relation_results}
\noindent
\textbf{\small Relationship among the variables}:

We have verified both the Pearson (linear) and the rank (non-parametric) correlation values between the variables population, LR and WPR 
in terms of the estimated parameters and the UP values for the years 2001 and 2011 separately;
the rank correlations are illustrated in the supplementary Figure \ref{FIG:Corr} (Appendix \ref{SEC:App}).
Interestingly, none of these correlations are significant at 95\% level; the same is also observed for the Pearson correlations.
This leads to the inference that the distributional uncertainty among these three indicators are mostly uncorrelated with each other
across all the states of India. It is an important observation in contrast to the fact that 
the distribution of the numbers of literate and working persons in a population depends directly (linearly) on 
the corresponding population distribution.

\subsection{Gender based analyses} 
\label{SEC:gender_results}

  The study on gender composition largely reflect the underlying social, economic and cultural patterns of the society in different ways. According to United Nation estimates, the world had 986 females against 1000 males in 2000. Interestingly the sheer weight of the population of the Asian countries, particularly China (944) and India (933) with low sex ratio contributes to the preponderance of males over females in world.

As per the Indian Census 2011, total population of India %is 1,21,01,93,422 which 
comprises of 62,37,24,248 males and 58,64,69,174 females with the sex ratio of 940 females per 1000 males. As per Census 2011, top five states/Union territories which have the highest sex ratio are Kerela (1,084) followed by Puducherry (1,038), Tamil Nadu (995), Andhra Pradesh (992) and Chhattisgarh (991). Five states which have the lowest sex ratio are Daman  Diu (618), Dadra  Nagar Haveli (775), Chandigarh (818), NCT of Delhi (866) and Andaman  Nicobar Islands (878). The  gender statistics revealed by Census India 2001, 532 million constituting 52 percent are males and 497 million constituting remaining 48 percent are females in the population.  Eighteen states/Uts have recorded sex ratio above the national average of 933, while remaining seventeen falls below this. Chandigarh and Daman  Diu occupy the bottom positions with less than 800 females per 1000 males. Though the Census has shown an increase in the sex ratio of total population from %927 in 1991 to 
933 in 2001 to 940 in 2011,  it still requires further improvement for the balance of male -female ratio.

In any country, higher literacy rates improve development indicators consistently.  In this respect, it is important to know the female percentage of the literate persons, as that may lead to better attainment of health and nutritional status, economic growth of community as a whole. The literacy rate of India in 2011 is 74.0 per cent. Literacy rate among females is 65.5 per cent whereas the literacy rate among males is 82.1 per cent showing a gap of 16.6 percent. If we see the data of  2001 census, the gap was 21.6 revealing development in this index .% and if we see the data of 1991 census, the gap was 24.8 percent. 
%Though the data reveals development in this index in last 2 census data, but much more has to be done for reduction of the gap.

Moreover, the male and female working population and work participation rate highlights the gender biased  occupational distribution of a region. %The temporal analysis  of total workers of India explains that the work participation rate has registered continuous increase in the last four decades. 
The Work Participation Rate (WPR), which is defined as the percentage of total workers to the total population, is 39.8 and  39.3 %and 37.5  
per cent as per the 2011 and  2001 %and 1991  
census. % and 39.3 percent in 2001, and 37.5 percent in 1991. 
 While the WPR for males marginally increased from 51.6 percent to 51.9 percent during 1991-2001, for females it improved significantly from 22.7 to 25.7 during the corresponding period. Interestingly, this increase is mainly due to increase in proportion of marginal workers which registered significant increase from 3.4 percent to 8.7 percent. According to  2001 census, in the total population percentage of male non-workers and female non-workers are given by 48.1 percent and  74.3 percent respectively. %The corresponding data for the year 1991 is given by  48.4 and 77.7. 

%XXXXXXXXXXXXXXXXx***************

This gender based information of the census data inspired us to perform a gender-based analysis of  the distribution of 
important socio-economic variables, such as population, literacy and work-participation rates of different states across India. 
According to our expectation, the analysis also revealed that both the male and female distributions fit very well with the DGBD. 
Besides, the DGB distribution fits very well with the male-female (sex) ratio of LR and WPR data for all the districts within  the states of India with a very low value of KS (RO).  For this purpose, we have computed the Sex-Ratio for the LR and WPR as follow:
\begin{eqnarray}
\mbox{Sex-Ratio of LR (SR-LR)} &=& \frac{\mbox{number of female literate in a district}}{\mbox{number of male literate  in that district}}\times 100,
\nonumber\\
\mbox{Sex-Ratio of WPR (SR-WPR)} &=& \frac{\mbox{number of female workers in a district}}{\mbox{number of male workers  in that district}}\times100.
\nonumber
\end{eqnarray}

\begin{figure}[!b]
	\centering
	\subfloat[SR-LR]{
	\includegraphics[width=0.45\textwidth]{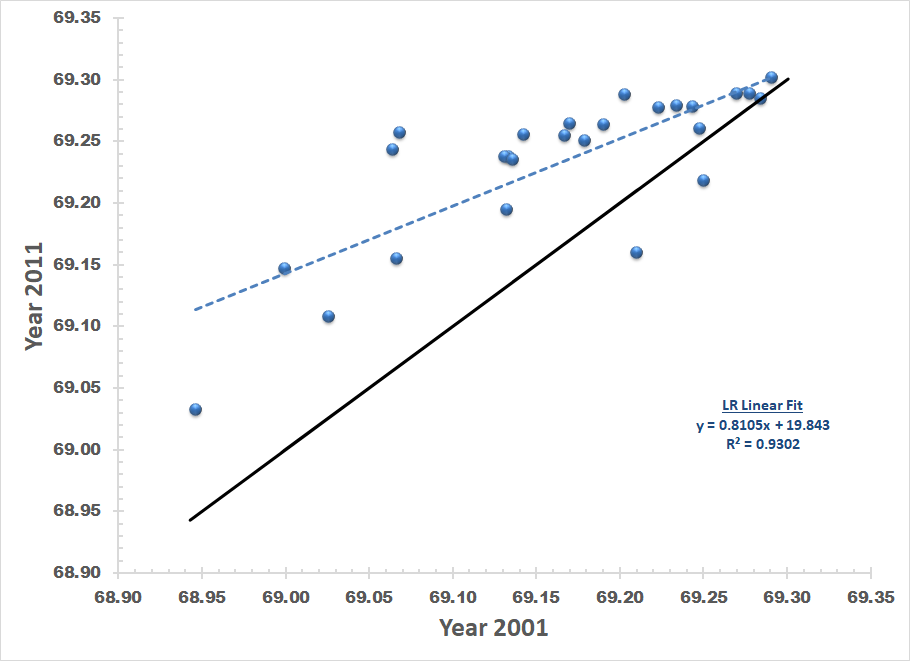}
	\label{FIG:loglogCumHaz_Veteran}}
	~ %--------------------------------------------------------------------
	\subfloat[SR-WPR]{
	\includegraphics[width=0.45\textwidth]{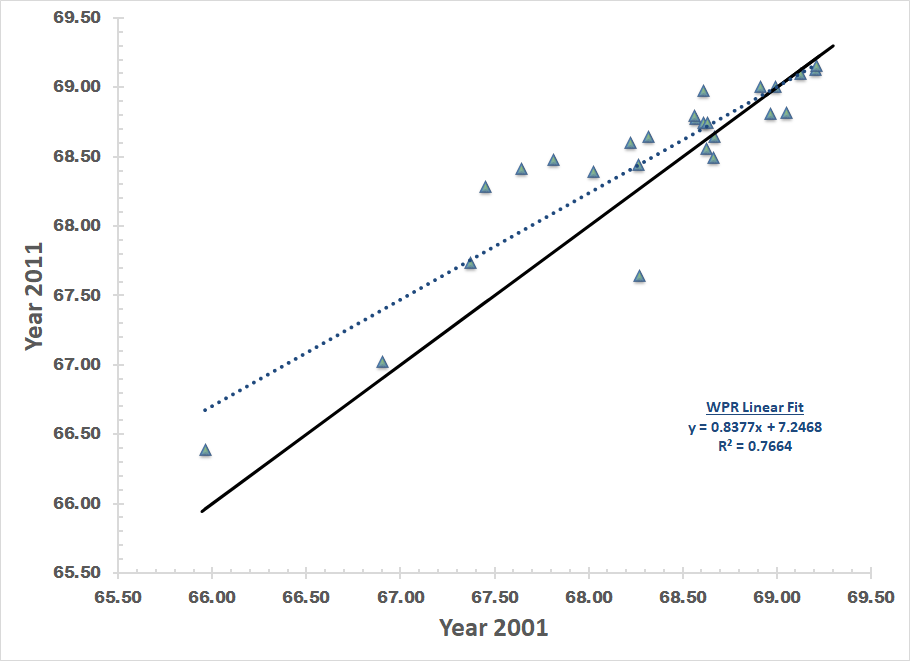}
		\label{FIG:loglogCumHaz_Veteran}}
	%~	%---------------------------------------------------------------------------
	\caption{Plot of the UP values for the gender based SR-LR and SR-WPR for different states in the years 2011 vs. 2001, 
		along with the corresponding linear fits. The black solid line represent the $y=x$ line (linear growth)}
	\label{FIG:SR_UP}
\end{figure}

\begin{table}[!h]
	\caption{The parameter estimates and the KS measures for %for 
	different states along
		 with total numbers (N) of districts and the UP values for Gender-based Analysis}
	%(***Need to change the Entropy values by UP values *****)}
	%	\resizebox{0.6\textwidth}{!}{ 
	\begin{tabular}{l|r|rrrr|rrrr}\hline
		&\multicolumn{1}{|c|}{}  &\multicolumn{4}{|c|}{SR-LR} &\multicolumn{4}{|c|}{SR-WPR} 	\\								
		States   &	$N$	 &$\widehat{a}$ 	&$\widehat{b}$	&KS&   UP &	$\widehat{a}$	&	$\widehat{b}$&	KS	& 	UP	\\
		\hline\hline
\multicolumn{10}{c}{\underline{Year 2011}}\\		
		Andhra Pradesh	&	23	&	0.079	&	0.027	&	0.003	&	69.24	&	0.010	&	0.292	&	0.004	&	68.77 \\	
		Arunachal Pradesh	&	16	&	-0.003	&	0.211	&	0.002	&	69.03	&	0.146	&	0.200	&	0.008	&	68.56\\	
		Assam	&	27	&	0.027	&	0.018	&	0.001	&	69.30	&	0.125	&	0.255	&	0.022	&	68.48\\	
		Bihar	&	38	&	0.031	&	0.036	&	0.001	&	69.29	&	0.178	&	0.162	&	0.010	&	68.60\\	
		Chhattisgarh	&	18	&	0.017	&	0.079	&	0.001	&69.26	&	0.028	&	0.157	&	0.006	&	69.10\\	
		Delhi	&	9	&	0.052	&	0.007	&	0.001	&	69.29	&	0.145	&	0.174	&	0.012	&	68.65\\	
		Gujarat	&	26	&	0.026	&	0.064	&	0.002	&	69.26	&	0.294	&	0.208	&	0.012	&	67.64\\	
		Haryana	&	21	&	-0.007	&	0.126	&	0.005	&	69.22	&	0.106	&	0.292	&	0.012	&	68.39\\	
		Himachal Pradesh	&	12	&	0.090	&	0.068	&	0.002	&	69.15	&	0.133	&	0.144	&	0.006	&	68.81\\	
		Jammu Kashmir	&	22	&	0.075	&	0.084	&	0.002	&	69.15	&	0.130	&	0.407	&	0.009	&	67.74\\	
		Jharkhand	&	24	&	0.051	&	0.044	&	0.001	&	69.26	&	0.062	&	0.332	&	0.007	&	68.44\\	
		Karnataka	&	30	&	0.064	&	0.046	&	0.001	&	69.24	&	0.065	&	0.162	&	0.003	&	69.01\\	
		Kerala	&	14	&	0.023	&	0.046	&	0.002	&	69.28	&	0.039	&	0.292	&	0.014	&	68.65\\	
		Madhya Pradesh	&	50	&	0.046	&	0.059	&	0.001	&	69.25	&	0.073	&	0.329	&	0.006	&	68.50\\	
		Maharashtra	&	35	&	0.049	&	0.026	&	0.000	&	69.28	&	0.027	&	0.294	&	0.003	&	68.74\\	
		Manipur	&	9	&	0.065	&	0.025	&	0.005	&	69.26	&	0.056	&	0.111	&	0.005	&	69.13\\	
		Meghalaya	&	7	&	0.153	&	0.015	&	0.001	&	69.11	&	0.257	&	-0.005	&	0.002	&	68.82\\	
		Mizoram	&	8	&	-0.014	&	0.163	&	0.003	&	69.16	&	0.114	&	0.035	&	0.004	&	69.16\\	
		Nagaland	&	11	&	0.038	&	0.037	&	0.002	&	69.28	&	-0.086	&	0.297	&	0.006	&	68.98\\	
		Odisha	&	30	&	0.046	&	0.062	&	0.002	&	69.24	&	0.023	&	0.702	&	0.014	&	67.03\\	
		Punjab	&	20	&	0.038	&	0.024	&	0.001	&	69.26	&	0.245	&	0.006	&	0.011	&	68.80\\	
		Rajasthan	&	33	&	0.034	&	0.081	&	0.001	&	69.24	&	0.086	&	0.141	&	0.003	&	69.01\\	
		Tamilnadu	&	32	&	0.047	&	0.030	&	0.001	&	69.28	&	0.117	&	0.192	&	0.007	&	68.75\\	
		Uttarakhand	&	13	&	0.067	&	0.068	&	0.002	&	69.19	&	-0.134	&	0.924	&	0.012	&	66.39 \\	
		Uttar Pradesh	&	71	&	0.047	&	0.047	&	0.001	&	69.26	&	0.142	&	0.314	&	0.019	&	68.28\\	
		West Bengal	&	19	&	-0.009	&	0.100	&	0.004	&	69.25	&	0.180	&	0.197	&	0.008	&	68.41\\	
%		\hline
%	\end{tabular}
%	\begin{tabular}{r|r|rrrr|rrrr}\hline
%	&\multicolumn{1}{|c|}{}  &\multicolumn{4}{|c|}{LR} &\multicolumn{4}{|c|}{WPR} 	\\								
%		STATES   &	$N$	 &$\widehat{a}$ 	&$\widehat{b}$	&KS&   Entropy&	$\widehat{a}$	&	$\widehat{b}$&	KS	& 	Entropy	\\\hline
		\hline\hline
\multicolumn{10}{c}{\underline{Year 2001}}\\				
		Andhra Pradesh	&	23	&	0.117	&	0.045	&	0.003	&	69.13	&	-0.027	&	0.387	&	0.005	&	68.57 \\
		Arunachal Pradesh	&	13	&	0.038	&	0.203	&	0.006	&	68.95	&	0.101	&	0.231	&	0.003	&	68.63\\
		Assam	&	23	&	0.026	&	0.037	&	0.003	&	69.29	&	0.193	&	0.314	&	0.014	&	67.81\\
		Bihar	&	37	&	0.045	&	0.091	&	0.002	&	69.20	&	0.153	&	0.292	&	0.014	&	68.22\\
		Chhattisgarh	&	16	&	0.089	&	0.073	&	0.005	&	69.14	&	0.056	&	0.116	&	0.008	&	69.13\\
		Delhi	&	9	&	0.052	&	0.029	&	0.002	&	69.27	&	0.164	&	0.148	&	0.009	&	68.67 \\
		Gujarat	&	25	&	0.022	&	0.118	&	0.002	&	69.19	&	0.186	&	0.227	&	0.004	&	68.27\\
		Haryana	&	19	&	0.063	&	0.036	&	0.001	&	69.25	&	0.041	&	0.445	&	0.007	&	68.02\\
		Himachal Pradesh	&	12	&	0.134	&	0.082	&	0.005	&	69.00	&	0.166	&	0.054	&	0.002	&	68.97 \\
		Jammu Kashmir	&	14	&	0.112	&	0.081	&	0.003	&	69.07	&	0.140	&	0.446	&	0.012	&	67.37 \\
		Jharkhand	&	18	&	0.097	&	0.099	&	0.002	&	69.07	&	0.041	&	0.389	&	0.014	&	68.27\\
		Karnataka	&	27	&	0.104	&	0.064	&	0.002	&	69.13	&	0.034	&	0.196	&	0.004	&	69.00\\
		Kerala	&	14	&	0.022	&	0.047	&	0.001	&	69.28	&	0.037	&	0.374	&	0.006	&	68.32\\
		Madhya Pradesh	&	45	&	0.055	&	0.098	&	0.002	&	69.18	&	0.070	&	0.280	&	0.006	&	68.66\\
		Maharashtra	&	35	&	0.067	&	0.047	&	0.001	&	69.23	&	-0.022	&	0.381	&	0.012	&	68.61\\
		Manipur	&	9	&	0.043	&	0.057	&	0.002	&	69.25	&	0.051	&	0.076	&	0.005	&	69.21\\
		Meghalaya	&	7	&	0.142	&	0.063	&	0.003	&	69.03	&	-0.010	&	0.208	&	0.002	&	69.05\\
		Mizoram	&	8	&	0.006	&	0.118	&	0.002	&	69.21	&	0.096	&	0.023	&	0.003	&	69.21\\
		Nagaland	&	8	&	0.065	&	0.053	&	0.002	&	69.22	&	-0.172	&	0.480	&	0.020	&	68.61\\
		Odisha	&	30	&	0.057	&	0.148	&	0.007	&	69.06	&	0.029	&	0.719	&	0.023	&	66.90\\
		Punjab	&	17	&	0.038	&	0.037	&	0.001	&	69.28	&	0.172	&	0.170	&	0.005	&	68.56\\
		Rajasthan	&	32	&	0.095	&	0.073	&	0.006	&	69.14	&	0.072	&	0.190	&	0.003	&	68.91\\
		Tamilnadu	&	30	&	0.068	&	0.036	&	0.002	&	69.24	&	0.027	&	0.323	&	0.005	&	68.63\\
		Uttarakhand	&	13	&	0.034	&	0.134	&	0.003	&	69.13	&	-0.185	&	1.054	&	0.020	&	69.13\\
		Uttar Pradesh	&	70	&	0.061	&	0.101	&	0.002	&	69.17	&	0.098	&	0.578	&	0.022	&	67.45\\
		West Bengal	&	18	&	-0.015	&	0.161	&	0.004	&	69.17	&	0.181	&	0.354	&	0.008	&	67.64\\
		\hline
	\end{tabular}
	%}
	\label{TAB:Gender_results}
\end{table}

\noindent
We apply our proposed analysis methodology for these two variables, depicting the gender based inequalities in LR and WPR, 
and the resulting parameter estimates and the UP values are presented in Table \ref{TAB:Gender_results}.
For better understanding of the change in distributional uncertainty from the year 2001 to 2011, 
the UP values for both SR-LR and SR-WPR are also presented graphically in Figure \ref{FIG:SR_UP} along with the linear fits. 
It can be seen that, on an average, there are sublinear growth of the distributional uncertainty (UP values) for both SR-LR and SR-WPR
between years 2001 and 2011, indicating reduced gender-discriminations in literacy rate as well as in work-participation rate.
For SR-LR, all states indeed have increased UP value in 2011, except for Haryana and Mizoram.
The lowest UP value (most discrimination) is observed in Arunachal Pradesh for SR-LR and in Uttarakhand for SR-WPR.
In general, the distributional uncertainty for SR-WPR remains mostly stable in the states having higher UP values,
with slight uniform variations.

%\newpage
%\subsection{Classification of States: How Different are they? }  		
%\label{SEC:classification}

\section{Conclusion}
% \begin{itemize}

In this paper we have quantified  the distributional uncertainty present in the various socio-economic indicators of all the Indian states, by studying the distribution of the districts within each individual states. The analysis is based on the DGB distribution, a typical rank-order distribution, which fits very well across many disciplines of Arts and Science.  
Despite the immense diversity of human settlements and extraordinary geographic variability, we have shown that all states obey DGB distribution not only for population size but also for other socioeconomic factors like literacy rate and work participation rate. %unemployment.
 The two model parameters of the distribution for different states construct a cluster. With the evolution of time there is no noted change in this cluster structure. 
Our primary analytical focus here is  concerned with the entropy of the distribution. Although, the value of the entropy for a particular state for all factors are different, the Uncertainty Percentage (UP), a measure defined for the uncertainty, is almost the same for all the states of different size, culture and economic and social conditions. This is one of the most intriguing outcome of this analysis. The results also indicate that this value of UP is also nearly equal to the value of the UP when it is studied with all the districts across the states. Remarkably, the study for different years also indicate a similar result.

%e analysis is that the relative entropy for each socioeconomic factors appear in a straight line .This behavior suggest that there is a universal social dynamics that inextricably linking this. (UNIVERSAL SOCIAL DYNAMICS$?$) 
 % \item uncertainty in entropy
  	%\end{itemize} 
		
It is important to note that the proposed UP measure of distributional uncertainty is quite a general formulation
which can further be applied to analyze the distributions of different important socio-economic variables 
of any country having stratified administrative structure.
Beside such applications, it would also be important to investigate the detailed theoretical properties of the proposed UP measure 
as possible future research works.

\bigskip \noindent
{\bf{Acknowledgment}}:\\
OM is thankful to Indian Statistical Institute, Kolkata, India for allowing her to work as a Visiting Student  
at the Physics and Applied Mathematics Unit of the Institute. SC thanks Science and Engineering Research Board (SERB), Government of India for financial support through Junior Research Fellowship (Grant No.CRG/2019/001461). The research of AG and BB is partially supported by a research grant (No. CRG/2019/001461) from the Science and Engineering Research Board (SERB), Government of India.

%XXXXXXXXXXXXXXXXXXXXXXXXXXXXXXXXXXXXXXXXXXXXXXXXXXXXXXXX

%\end{document}

%XXXXXXXXXXXXXXXXXXXXXXXXXXXXXXXXXXXXXXXXXXXXXXXXXXXXXXXXXXXXXXxxx

\newpage
\appendix
\section{Supplementary Tables and Figures}
\label{SEC:App}

\setcounter{table}{0}
\renewcommand{\thetable}{S\arabic{table}}
\setcounter{figure}{0}
\renewcommand{\thefigure}{S\arabic{figure}}

\begin{table}[h]
\caption{Population (Pop), Literacy Rate (LR) and Work Participation Rate (WPR), along with the number of districts ($N$), for every states of India} 
	\centering	
	%	\resizebox{0.6\textwidth}{!}{
	\begin{tabular}{l|rrrr|rrrr}\hline
		%& \multicolumn{1}{|c|}{States} 	& \multicolumn{3}{|c|}{2011} 	& \multicolumn{3}{|c|}{2001} 	\\ 
		&\multicolumn{4}{|c|}{2011} &\multicolumn{4}{|c}{2001}\\
	State	&	$N$ &  Pop 	&	LR	&	WPR	 &  $N$ &  Pop	&	LR	&	WPR \\ \hline
	Andaman Nicobar (UT)	& 3& 	380581	&	77.32	&	40.08	& 2 & 	356152	&	71.07	&	38.26\\
	Andhra Pradesh	& 23 &	84580777	&	59.77	&	46.61	& 23&	76210007	&	52.40	&	45.79\\
	Arunachal Pradesh	& 16 &	1383727	&	55.36	&	42.47	&13&	1097968	&	44.15	&	43.98\\
	Assam	&	27 & 31205576	&	61.46	&	38.36	&	23&26655528	&	52.58	&	35.78\\
	Bihar	&	38 & 104099452	&	50.44	&	33.36	&37&	82998509	&	37.48	&	33.70\\
	Chandigarh (UT)	& 1&	1055450	&	76.31	&	38.29	& 1&	900635	&	71.42	&	37.80\\
	Chattisgarh	& 18& 	25545198	&	60.21	&	47.68	&16&	20833803	&	53.63	&	46.46\\
Dadra - Nagar Haveli (UT)	& 1&	343709	&	64.95	&	45.73	&1&	220490	&	47.12	&	51.76\\
	Daman - Diu (UT)	& 2&	243247	&	77.45	&	49.86	&2&	158204	&	68.01	&	46.01\\
	Delhi (UT)	&	9& 16787941	&	75.87	&	33.28	&	9& 13850507	&	69.78	&	32.82\\
	Goa	& 2&	1458545	&	79.91	&	39.58	& 2& 	1347668	&	73.13	&	38.80\\
	Gujarat	& 26& 	60439692	&	67.99	&	40.98	&	25& 50671017	&	58.87	&	41.95\\
	Haryana	& 21&	25351462	&	65.48	&	35.17	&	19&21144564	&	57.20	&	39.62\\
	Himachal Pradesh	& 12&	6864602	&	73.42	&	51.85	&12 &	6077900	&	66.50	&	49.24\\
	Jammu Kashmir	& 22& 12541302	&	56.35	&	34.47	& 14&	10143700	&	47.39	&	37.01\\
	Jharkhand	& 24& 32988134	&	55.56	&	39.71	& 18& 	26945829	&	43.71	&	37.52\\
	Karnataka	& 30&	61095297	&	66.53	&	45.62	& 27&	52850562	&	57.59	&	44.53\\
	Kerala	&14&	33406061	&	84.22	&	34.78	&14&	31841374	&	80.04	&	32.30\\
	Lakshadweep (UT)	& 1&	64473	&	81.51	&	29.09	&	1& 60650	&	73.67	&	25.32\\
	Madhya Pradesh	&	50&72626809	&	59.00	&	43.47	& 50&	60348023	&	52.35	&	42.74\\
	Maharashtra	&	35 &112374333	 &	72.57	&	43.99	&35&	96878627	&	66.03	&	42.50\\
	Manipur	&	9& 2855794	&	66.83	&	45.68	&9&	2293896	&	57.13	&	41.21\\
	Meghalaya	& 7& 	2966889	&	60.16	&	39.96	& 7& 	2318822	&	49.93	&	41.84\\
	Mizoram	&	8& 1097206	&	77.30	&	44.36	& 8&	888573	&	74.44	&	52.57\\
	Nagaland	&	11 & 1978502	&	67.85	&	49.24	& 8& 	1990036	&	56.90	&	42.60\\
	Odisha	& 30& 	41974218	&	63.71	&	41.79	& 30&	36804660	&	53.90	&	38.79\\
	Puducherry (UT)	& 4&	1247953	&	76.71	&	35.66	& 4&	974345	&	71.47	&	35.17\\
	Punjab	&	20& 27743338	&	67.43	&	35.67	& 17&	24358999	&	60.58	&	37.47\\
	Rajasthan	&	33& 68548437	&	55.84	&	43.60	& 32&	56507188	&	49.02	&	42.06\\
	Sikkim	&	4& 610577	&	72.87	&	50.47	& 4&	540851	&	58.86	&	48.64\\
	Tamilnadu	&	32& 72147030	&	71.85	&	45.58	&30&	62405679	&	64.94	&	44.67\\
	Tripura	& 4&	3673917	&	76.34	&	40.00	&4&	3199203	&	63.21	&	36.25\\
	Uttar Pradesh	& 71& 	199812341	&	57.25	&	32.94	& 70 &	166197921	&	45.56	&	32.48\\
	Uttarakhand	& 13&	10086292	&	68.22	&	38.39	& 13&	8489349	&	60.14	&	36.92\\
	West Bengal	& 19&	91276115	&	67.42	&	38.08	& 18&	80176197	&	58.87	&	36.77\\
		\hline
	\end{tabular}
	%}
	\label{TAB:Primary_data}
\end{table}
%*P= Total Population of a state, LR= Literacy rate, WPR= Work Partitipation Ratio

%\begin{figure}[!h]
%	\centering
%	\subfloat[Year 2011]{
%		\includegraphics[width=0.33\textwidth]{CorrRank_2011_a.png}
%		\includegraphics[width=0.33\textwidth]{CorrRank_2011_b.png}
%		\includegraphics[width=0.33\textwidth]{CorrRank_2011_UP.png}
%		\label{FIG:loglogCumHaz_Veteran}}
%	\\ %--------------------------------------------------------------------
%	\subfloat[Year 2001]{
%		\includegraphics[width=0.33\textwidth]{CorrRank_2001_a.png}
%		\includegraphics[width=0.33\textwidth]{CorrRank_2001_b.png}
%		\includegraphics[width=0.33\textwidth]{CorrRank_2001_UP.png}
%		\label{FIG:loglogCumHaz_SmallCell}}
%	%---------------------------------------------------------------------------
%	\caption{Spearman Rank Correlation of the estimated parameters and UP values for Populations, LR and WPR in 2011 and 2001}
%	\label{FIG:Corr}
%\end{figure}

\begin{figure}[!h]
\begin{minipage}{0.48\textwidth}
	\centering
\subfloat[Year 2011]{
\includegraphics[width=0.99\textwidth]{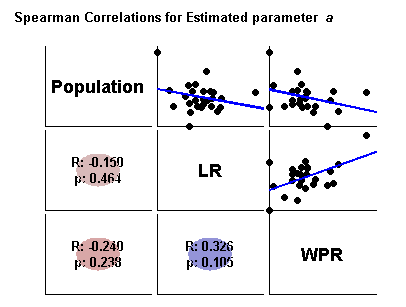}
}\\
\subfloat[Year 2011]{
\includegraphics[width=0.99\textwidth]{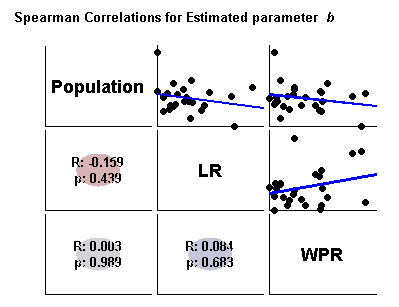}
}\\
\subfloat[Year 2011]{
\includegraphics[width=0.99\textwidth]{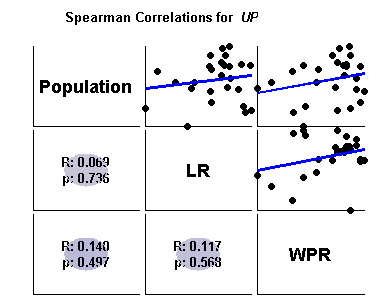}}
%---------------------------------------------------------------------------
\end{minipage}\hfill % maximize space between the minipages
\begin{minipage}{0.48\textwidth}
	\centering
\subfloat[Year 2001]{
\includegraphics[width=0.99\textwidth]{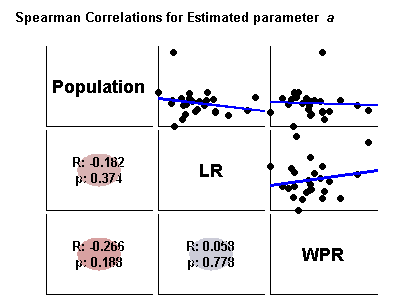}
}\\
\subfloat[Year 2001]{
\includegraphics[width=0.99\textwidth]{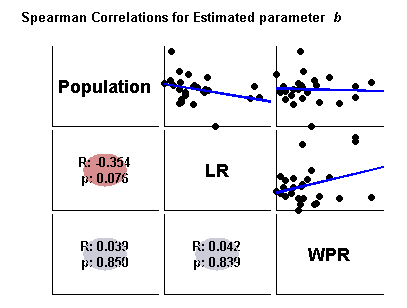}
}\\
\subfloat[Year 2001]{
\includegraphics[width=0.99\textwidth]{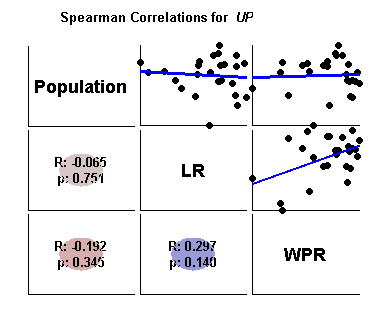}}
%---------------------------------------------------------------------------
\end{minipage}
\caption{Spearman Rank Correlation of the estimated parameters and UP values for Populations, LR and WPR in 2011 and 2001}
\label{FIG:Corr}
\end{figure}

 	~ %--------------------------------------------------------------------
 	%	\subfloat[2017]{
 	%	\includegraphics[width=0.47\textwidth]{USA_POPESTIMATE2017.png}
 	%	\label{FIG:loglogCumHaz_SmallCell}}
 	%---------------------------------------------------------------------------
 	%---------------------------------------------------------------------------
 	%\caption{Entropy plot  for the three quantities for each state of India 2011}
 	
 	%\label{FIG:2011}
 %\end{figure}

 %\begin{figure}[!h]
	%\centering
	%\subfloat[2001]{
		%\includegraphics[width=0.57\textwidth]{ENTROPY PLOT01.png}
		%\label{FIG:}}
	~ %--------------------------------------------------------------------
	%	\subfloat[2017]{
	%	\includegraphics[width=0.47\textwidth]{USA_POPESTIMATE2017.png}
	%	\label{FIG:loglogCumHaz_SmallCell}}
	%---------------------------------------------------------------------------
	%---------------------------------------------------------------------------
	%\caption{Entropy plot for the three quantities for each state of India 2011}
	
	%\label{FIG:2001}
%\end{figure}

%\begin{figure}[!h]
	%\centering
	%\subfloat[1991]{
		%\includegraphics[width=0.57\textwidth]{ENTROPY PLOT91.png}
		%\label{FIG:loglogCumHaz_Veteran}}
	~ %--------------------------------------------------------------------
	%	\subfloat[2017]{
	%	\includegraphics[width=0.47\textwidth]{USA_POPESTIMATE2017.png}
	%	\label{FIG:loglogCumHaz_SmallCell}}
	%---------------------------------------------------------------------------
	%---------------------------------------------------------------------------
	%\caption{Entropy plot for the three quantities for each state of India 1991}
	
	%\label{FIG:1991}
%\end{figure}

%\begin{figure}[!h]
	%\centering
	%\subfloat[2011]{
		%\includegraphics[width=0.57\textwidth]{Plot_ab11.png}
	%	\label{FIG:loglogCumHaz_Veteran}}
	~ %--------------------------------------------------------------------
	%	\subfloat[2017]{
	%	\includegraphics[width=0.47\textwidth]{USA_POPESTIMATE2017.png}
	%	\label{FIG:loglogCumHaz_SmallCell}}
	%---------------------------------------------------------------------------
	%---------------------------------------------------------------------------
	%\caption{Plot of the estimated parameters $a,b$ for the three quantities for each state of India 2011}
	
	%\label{FIG:2011}
%\end{figure}

%\begin{figure}[!h]
%	\centering
	%\subfloat[2001]{
		%\includegraphics[width=0.47\textwidth]{Plot_ab2001.png}
		%\label{FIG:loglogCumHaz_Veteran}}
	~ %--------------------------------------------------------------------
	%	\subfloat[2017]{
	%	\includegraphics[width=0.47\textwidth]{USA_POPESTIMATE2017.png}
	%	\label{FIG:loglogCumHaz_SmallCell}}
	%---------------------------------------------------------------------------
	%---------------------------------------------------------------------------
	%\caption{Plot of the estimated parameters $a,b$ for the three quantities for each state of India 2001}
	
	%\label{FIG:2001}
%\end{figure} 

%\begin{figure}[!h]
	%\centering
	%\subfloat[1991]{
		%\includegraphics[width=0.47\textwidth]{Plot_ab1991.png}
	%	\label{FIG:loglogCumHaz_Veteran}}
	~ %--------------------------------------------------------------------

\end{document}